# 3DPX: Single Panoramic X-ray Analysis Guided by 3D Oral Structure Reconstruction

Xiaoshuang Li[1,2], Zimo Huang[2], Mingyuan Meng[1,2], Eduardo Delamare[2], Dagan Feng[2], Lei Bi[1,2]*, Bin Sheng[1]*, Lingyong Jiang[3]*, Bo Li[4]*, Jinman Kim[2]

[1] Department of Computer Science and Engineering, Shanghai Jiao Tong University, Shanghai 200240, China
[2] School of Computer Science, University of Sydney, Sydney NSW 2000, Australia
[3] Department of Oral and Cranio-maxillofacial Surgery, Shanghai Ninth People's Hospital,
Shanghai Jiao Tong University School of Medicine, Shanghai 200240, China
[4] State Key Laboratory of Oral & Maxillofacial Reconstruction and Regeneration,
Key Laboratory of Oral Biomedicine Ministry of Education, Hubei Key Laboratory of Stomatology,
School & Hospital of Stomatology, Wuhan University, Wuhan 430072, China

**Abstract**
Panoramic X-ray (PX) is a prevalent modality in dentistry practice owing to its wide availability and low cost. However, as a 2D projection of a 3D structure, PX suffers from anatomical information loss and PX diagnosis is limited compared to that with 3D imaging modalities. 2D-to-3D reconstruction methods have been explored for the ability to synthesize the absent 3D anatomical information from 2D PX for use in PX image analysis. However, there are challenges in leveraging such 3D synthesized reconstructions. First, inferring 3D depth from 2D images remains a challenging task with limited accuracy. The second challenge is the joint analysis of 2D PX with its 3D synthesized counterpart, with the aim to maximize the 2D-3D synergy while minimizing the errors arising from the synthesized image. In this study, we propose a new method termed 3DPX – PX image analysis guided by 2D-to-3D reconstruction, to overcome these challenges. 3DPX consists of (i) a novel progressive reconstruction network to improve 2D-to-3D reconstruction and, (ii) a contrastive-guided bidirectional multimodality alignment module for 3D-guided 2D PX classification and segmentation tasks. The reconstruction network progressively reconstructs 3D images with knowledge imposed on the intermediate reconstructions at multiple pyramid levels and incorporates Multilayer Perceptrons (MLPs) to improve semantic understanding. The downstream networks leverage the reconstructed images as 3D anatomical guidance to the PX analysis through feature alignment, which increases the 2D-3D synergy with bidirectional feature projection and decease the impact of potential errors with contrastive guidance. Extensive experiments on two oral datasets involving 464 studies demonstrate that 3DPX outperforms the state-of-the-art methods in various tasks including 2D-to-3D reconstruction, PX classification, and PX lesion segmentation. The robust performance and the well-structured pipeline of 3DPX suggest its potential applicability to general PX image analysis.

# Introduction

Panoramic X-ray (PX), an extra-oral imaging technique, is widely used in dental practices for diagnostic, assessment and, monitoring purposes (Katsumata, 2023; Różyło-Kalinowska, 2021; Shahidi et al., 2018). It generates stretched 2-dimensional (2D) images of the entire maxillomandibular area by rotating an X-ray emitter around the patient's head along a curved trajectory while capturing projections of the anatomical structures. When compared to other dental imaging modalities such as cone-beam computed tomography (CBCT), magnetic resonance imaging (MRI), and ultrasonography (US), PX is most routinely acquired and has the advantages in lower cost, patient convenience, and lower radiation dose (Song et al., 2021). However, as a flat projection image, PX lacks 3D anatomical information, which impedes accurate disease interpretation (Estrela et al., 2008; Izzetti et al., 2021; Tsai et al., 2012) in downstream tasks such as lesion segmentation, disease classification, and angular misalignment detection (Delamare et al., n.d.). In comparison, CBCT has higher specificity and excellent accuracy in dental image analysis (Karamifar, 2020).

* Corresponding authors.
E-mail addresses: lei.bi@sjtu.edu.cn (L. Bi), shengbin@sjtu.edu.cn (B. Sheng) jianglingyong@sjtu.edu.cn (L. Jiang) libocn@whu.edu.cn (B. Li)
Phone number: +86 18254612755 (L. Bi), +86 15026790946 (B. Sheng), (L. Jiang), (B. Li)

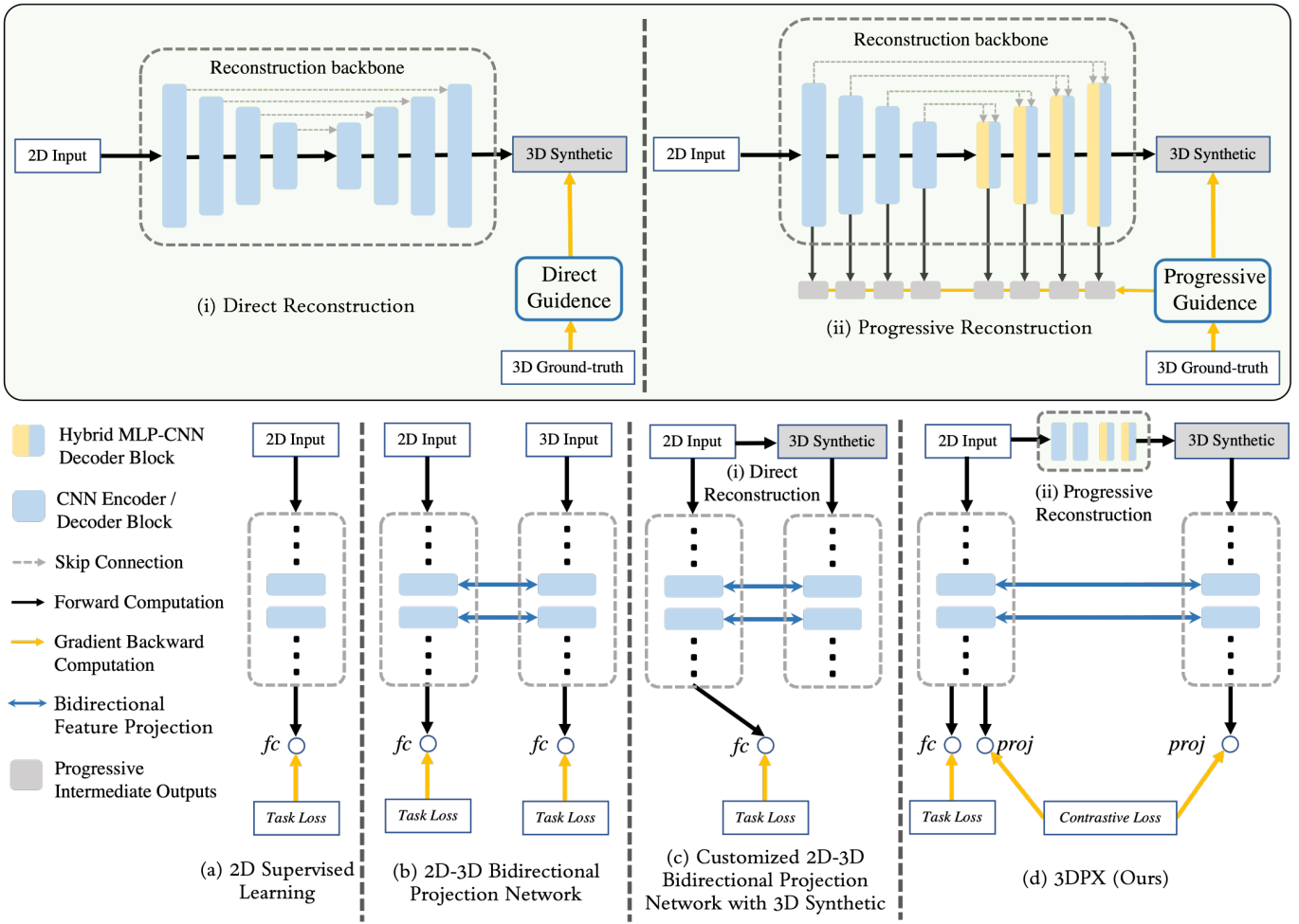

**Fig. 1.** A conceptual illustration of 3DPX for 2D-3D joint PX analysis. (a-d) presents the difference between: (a) 2D-based learning, (b) 2D-3D joint learning with 3D ground-truth, (c) 2D-3D joint learning without 3D ground-truth, and (d) our 3DPX with bidirectional contrastive loss. (i-ii) presents the difference between (i) a regular encoder-decoder direct reconstruction backbone used in (c) and (ii) our progressive reconstruction architecture that support the joint analysis of (d).

The motivation to reconstruct 3D spatial information from 2D PX has seen sustained research interests, with the aim to enable PX to retain the quality of diagnostic information and accuracy of pathology identification as seen on CBCT without the potential negative impacts of the latter. Song et al. (Song et al., 2021) developed Oral-3D, a Generative Adversarial Network (GAN) model with a residual Convolutional Neural Network (CNN) generator. Liang et al. (Liang et al., 2020) proposed a CNN-based architecture to firstly segment the PX images and then generate voxelized teeth based on the segmentation masks. These works use generative AI algorithms or traditional non-generative CNN networks to reconstruct 3D representations based on the training data. These works demonstrate the technical feasibility of 2D-to-3D dental structure reconstruction. However, there are two main challenges in leveraging the 3D reconstruction to enhance the analysis of 2D PX images.

First, existing 2D-to-3D reconstruction methods yield results with limited accuracy for PX images. These methods employ CNNs to directly map 2D PX images to 3D image volumes, as illustrated in Fig. 1(i). However, this simple mapping underestimates the complexity of 2D-to-3D reconstruction and cannot fully handle the difficulties in inferring depth-axis spatial information from 2D images with only height and width axes. While it's an intuitive solution to use the feature channel of 2D convolution output as the representation of 3D depth information, these CNN-based methods are limited by the intrinsic locality of convolutional operations. Further, the intermediate feature maps of existing methods are not fully leveraged such that the reconstruction results often lack details and tend to generate artefacts.

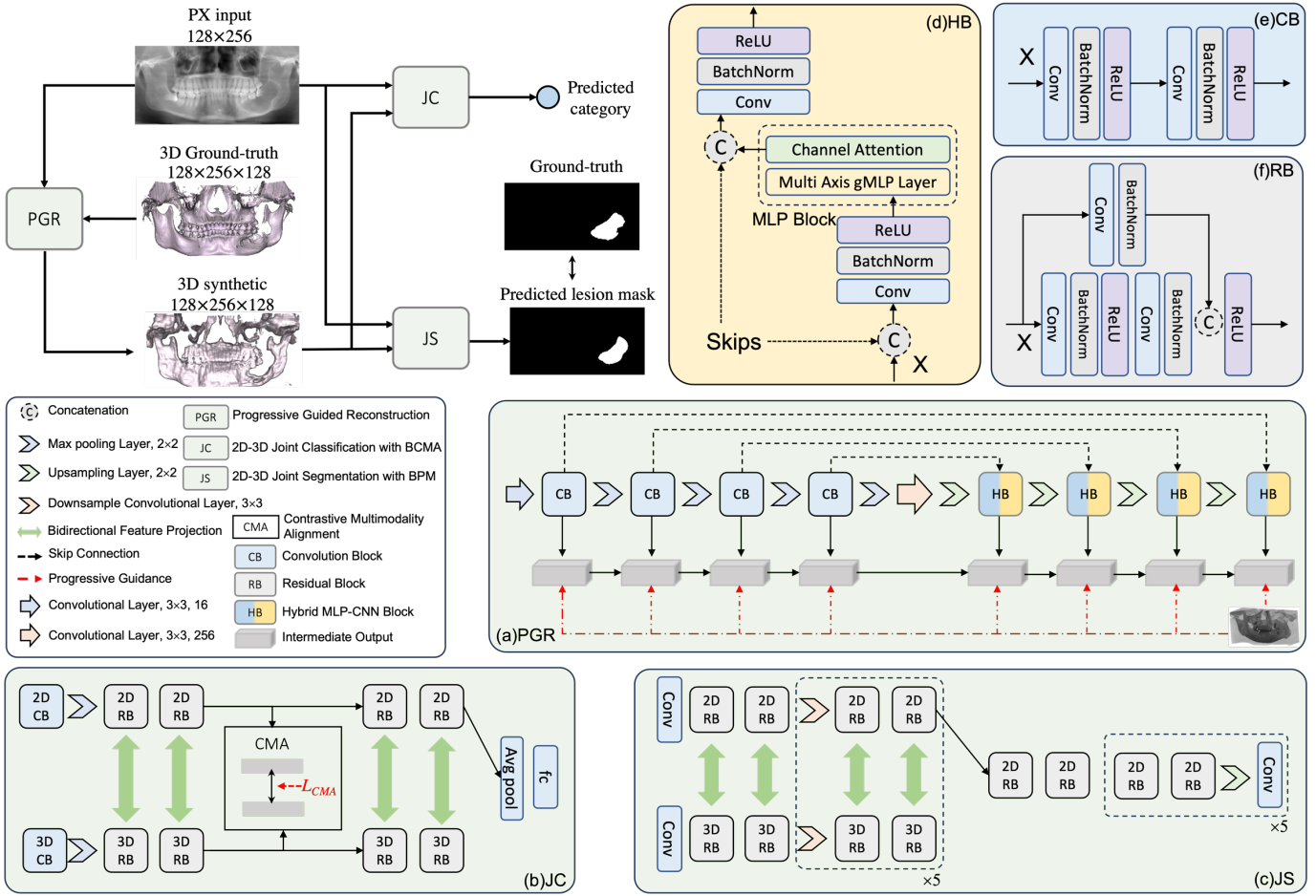

**Fig. 2.** The workflow of our 3DPX for 2D-3D joint PX analysis.

Second, joint analysis (or fusion) of 2D real and 3D synthesized images is also a challenging task. Existing 2D-3D fusion strategies were mainly designed for real captured 3D scene point cloud and its corresponding 2D images (Hu et al., 2021; Jaritz et al., 2019; Kweon and Yoon, n.d.). Compared to the synthetic data, these real captured data representations are of higher credibility, thus existing fusion methods directly involves available features via concatenation or other direct aggregation schemes (Fig. 1b). When applied to jointly analysis 3D reconstruction and 2D PX images, these methods did not take into consideration the inherent discrepancy between the synthetic and the real data. Meanwhile, applying existing methods to 2D-3D joint learning with 3D synthetic data faces a crucial drawback due to the absence of corresponding 3D ground-truth (Fig. 1c) on dense prediction tasks. This induces negative impacts when handling multimodality features fusion with the synthetic data, including the possible shortfall of synthetic information that are not fully restored from the PX images, and the potential error caused by the generation models.

In this study, we propose a new method termed 3DPX – PX image analysis guided by 2D-to-3D reconstruction, to improve both the 2D-to-3D reconstruction quality and the subsequent joint analysis of the real 2D and synthetic 3D images. A conceptual explanation of 3DPX's design is illustrated in Fig. 1d, and its workflow is depicted in Fig. 2. It firstly reconstructs 3D oral structure from PX images with the proposed progressive guided reconstruction (PGR) module, which incorporates a pyramid network consisting of convolutional blocks and Hybrid Multilayer Perceptron (MLP)-CNN Blocks (HB). MLPs have demonstrated strong capabilities in capturing fine-grained long-range dependence among high-resolution image details (Meng et al., 2023). HB integrates MLPs that captured fine-grained long-range dependency and CNNs to improve the semantic understanding during reconstruction. During this process,

3D images are progressively reconstructed with knowledge imposed on the intermediate reconstruction result at multiple pyramid levels. Subsequently, the reconstructed 3D structure and the real PX image together act as the input of the downstream networks, which employs bidirectional contrastive multimodality alignment (BCMA) to jointly give classification prediction on angular misalignment error of PX, or segmentation prediction of Odontogenic Cystic Lesions (OCLs). BCMA facilitates the feature fusion by bidirectional feature projection to further boosts the synergy and mitigates the feature representation discrepancy and uses contrastive penalty to constrain the latent distance between the synthetic and real images to minimum potential errors. Our method was comprehensively evaluated on three tasks and demonstrated superior capability on analyzing PX images. The main technical contributions of 3DPX are summarized as follows:

- We propose a progressive reconstruction strategy where the 3D images are reconstructed with progressive guidance imposed on the intermediate reconstruction results at each pyramid level, as illustrated in Fig. 1(ii), thus resulting in more fine-grained reconstruction.

- Our 3DPX integrates the advantages of MLPs and CNNs, such that it allows the capture of long-range visual dependence and small subtle details, and thus improving the semantic information during reconstruction.

- We propose a 2D-3D bidirectional feature projection without 3D labels, to conduct 2D-3D joint learning with 3D synthetic data, thus facilitating the synergy between the synthetic and real images and mitigating feature representation discrepancy.

- We introduce contrastive multimodality alignment as a constraint, which operate on the distance of latent space of the 3D synthetic and real images, thus mitigate the impact of potential errors on the synthetic data.

In our previous study published in a conference processing (Li et al., 2024), we reported our preliminary results which focused on 2D-to-3D reconstruction of oral PX images, and not in how the two modalities can be optimally combined for downstream tasks. In this study, we extended our preliminary study as follows: (1) the scope of the problem has been broadened from solely on 2D-3D reconstruction to a more developed objective: the joint analysis of real and synthetic images; (2) an introduction of BCMA module to handle the feature alignment between 2D real images and 3D synthetic images; (3) new data for the training and evaluation of 3DPX, from 464 patient data in (Li et al., 2024) to now involve 217 more OCLs segmentation labels, and includes OCLs segmentation tasks and, (4) extended comparisons to the state-of-the-art, detailed ablation study and, thorough discussions on our method's performance across multiple new downstream tasks.

## Related Work

### A. Deep Learning for Panoramic X-ray Analysis

Research in PX analysis with deep learning method covers a range of topics, including segmentation and detection of teeth and/or diseases. Zhu *et al*. proposed CariesNet (Zhu et al., 2023) to delineate three types of caries lesions. It integrated an encoder-decoder with a full-scale axial attention module to strengthen the network's attention towards caries lesions presented as small targets. Zhao *et al*. proposed TSASNet (Zhao et al., 2020), a two stage network for tooth segmentation using long short-term memory (LSTM) modules that encoded pixel-wise attention and a U-shape segmentation sub-network. Hamamci *et al*. (Hamamci et al., 2023) formulated the tooth detection problem as a denoising process of noisy boxes and proposed a diffusion-based multi-label tooth and disease detection framework. Their results reveal significant advancements in the diagnostic capabilities of PX analysis, demonstrating its broad applicability across various dental healthcare scenarios. However, despite these achievements, the existing methods were developed to address specific PX analysis problems and have not yet leveraged 3D imaging modalities for a more comprehensive diagnostic approach. The possible complementary information from 3D reconstruction that could contribute to various aspects of PX analysis remains unaddressed.

### B. 2D-to-3D Reconstruction from X-ray

State-of-the-art 2D-to-3D reconstruction from X-ray has numerous applications in multiple imaging modalities, such as X2CT-GAN for lung X-ray, Oral-3D for PX, and several other approaches. These methods can be categorized into two main categories of template deformation and CT reconstruction. Kim et al. (Kim et al., 2019) proposed a method based on modifiable leg bone template, which applied CNN-based feature analysis on X-Ray images to modified the preset parameters on the 3D template model. In dental area, Chen et al. (Chen et al., 2023) proposed to extract teeth contour from intra-oral imaging and iteratively fit the parametric teeth models to finally achieve 2D-to-3D teeth reconstruction. These template-based reconstruction methods focus on the morphology of specific structures and requires predefined shape and modifiable parameters. While the reconstructed outcomes emphasize shape and contour and are applicable for specific purposed, such as structure modeling, many contexts require CT reconstruction that recover more generalized 3D information.

Methods for CT reconstruction from X-ray primarily rely on using CNNs to directly predict 3D structures from 2D PX images using a 2D encoder-decoder network architecture. By using 2D convolutional layers on 3D image data, the depth information of 3D images is processed as the feature channels in the 2D convolutional layers. Ying et al. (Ying et al., 2019) introduced X2CT-GAN with a novel feature fusion method and the GAN framework to reconstruct CT from two orthogonal X-rays. Wang et al. (Wang et al., 2023) further extended it into TRCT-GAN structure with an additional transformer module. These methods leveraged two or more view perspectives to mitigate reconstruction artifacts and enrich multi-view information. In the realm of oral and maxillofacial radiology, Song et al. (Song et al., 2021) proposed Oral-3D, a GAN model with a Residual CNN generator, which was the first attempt for cross-dimension reconstruction of single PX images. Song et al. (Song et al., 2021) demonstrate that the 2D encoder-decoder structure can facilitate the 3D reconstruction for PX, which captured by rotational projection techniques compared to other directly projected X-rays. However, it shares the same limitation with other CNN-based method imposed by the intrinsic locality of convolution operations. Further, the intermediate feature maps are not fully leveraged such that the reconstruction results often lack details and tend to generate artefacts.

### C. 2D-3D Joint Learning

Methods designed for the integration of multimodality data representations are widely employed in medical imaging analysis to promote complementarity of information among imaging modalities, e.g., between Positron Emission Tomography (PET) and CT (Meng et al., 2022; Peng et al., 2019) and between multi-parametric Magnetic Resonance Imaging (MRI) (Chen et al., 2024). While PET captures metabolically function of tissues, CT provides precise details of structures, and MRI with different parameter settings focuses on different tissue properties such as water content or proton density. In combination, these heterogeneous features complement each other and facilitate imaging analysis. However, based on our review, there is an unexplored area in the development of joint learning methods designed for the integration of 2D-3D medical imaging pairs, such as with PX and CBCT pair.

2D-3D joint learning methods has been explored mainly in scene semantic segmentation using 2D natural images and 3D point cloud data acquired by Light Detection and Ranging (LiDAR) systems. They can be categorized into two feature fusion schemes: unidirectional (Hou et al., n.d.; Liu et al., n.d.) and bidirectional (Hu et al., 2021; Kweon and Yoon, n.d.) feature projection. Liu *et al.* (Liu et al., n.d.) proposed the use of 3D features extracted from large-scale point cloud data to improve segmentation on 2D natural images via knowledge distillation. The 2D feature extractor simulates the 3D features and predicts segmentation on the 3D scene, meanwhile the parameters were guided by the 3D feature extractor. Kweon *et al.* (Kweon and Yoon, n.d.) proposed cross-modality losses and used it to mutually refine the segmentation by projecting the network outputs between 3D point cloud and 2D image representations. Hu *et al.* proposed BPNet (Hu et al., 2021), another bidirectional projection method based on 2D-3D joint network designed for point cloud and images based on sparse 3D U-Net. In comparison, it exchanges features at multiple pyramid levels on the decoder branch of the U-Net structure. These methods build up correspondence between points in 3D scene and pixels in 2D image by rendering geometric constraints. Thus, it transmits and then fuses the features between two modalities. In all these methods, they were designed for real dataset. To leverage 2D-3D joint learning methods on paired PX and synthesized CBCT data, constraints need to be set during feature projection to improve fusion effectiveness and control potential error.

## Method

### A. Overview

Fig. 2 shows the workflow of the proposed 3DPX. It takes a 2D PX image (128×256) as input, reconstructs the corresponding unfolded 3D structure (128×256×128), and performs 2D-3D joint analysis for downstream PX classification and segmentation tasks. Our 3DPX consists of two main components: a progressive-guided reconstruction (PGR) backbone for 2D-to-3D PX reconstruction (refer to Section III-B and Fig. 2(a)) and a cascaded 2D-3D joint analysis network based on feature projection network (FPN) (Hu et al., 2021) for the downstream tasks (refer to Section III-C, Fig. 2(b) and (c)). The 3D structures reconstructed by PGR are fed into the 3D branch of joint classification or segmentation network, together with the original PX images which are fed into the parallel 2D branch, to conduct 2D-3D joint analysis.

## B. Progressive Guided Reconstruction (PRG)

### 1) Progressive reconstruction strategy

In this section, we introduce the progressive guidance strategy for 2D-to-3D PX reconstruction. The PGR backbone is a U-shaped structure consisting of 2D convolutional (Conv) blocks in the encoder and Hybrid MLP-CNN blocks in the decoder. The encoder starts with a convolutional layer that maps the single-channel input to a feature space with size of 16. The details of Conv block are shown in Fig. 2(e), which consists of two convolutional layers followed by Batch Normalization (BN) and ReLU activation. The encoder increases the size of depth channel to 128 and decrease the width and height to [16, 32]. The encoder and decoder are connected by a bottleneck convolutional layer that increases the feature channel size to 256.

Let B be a given encoder or decoder block in 3DPX, our PRG strategy introduces multiple guidance by applying penalties $\mathcal{L}$ on the intermediate output of B. Specifically, the penalty $\mathcal{L} = \mathcal{L}_{SSE}(f, X)$ where $f = B(X)$ denotes the intermediate feature map, Y denotes the scaled label, and $\mathcal{L}_{SSE}$ represents the Error Sum of Square (SSE) loss. At each reconstruction stage i, the guided penalty is formulated as:

$$\begin{aligned}\mathcal{L}_0(X) &= \mathcal{L}_{SSE}(B_0(X), Y_0) \\ \mathcal{L}_1(X) &= \mathcal{L}_{SSE}(B_1 \circ B_0(X), Y_1) \\ &\cdots \\ \mathcal{L}_i(X) &= \mathcal{L}_{SSE}(B_i \circ B_{i-1} \circ \cdots \circ B_0(X), Y_i).\end{aligned} \quad (1)$$

As intermediate reconstructions progressively improve and approach the final output, the intermediate penalties should accordingly have progressive weight to emphasize the guidance close to the output layer and downplay the role of guidance close to the input layer. To achieve this, a set of hyper-parameters α is set on $L_i(X)$ and the final training loss function for PRG $L_{PR}$ is formulated as

$$\mathcal{L}_{PR} = \sum_{i=0}^{n-1} \alpha_i \cdot \mathcal{L}_i(X). \quad (2)$$

where n is the number of the encoder and decoder blocks. In the experiment, we set $\alpha_i = 2^{n-1-i}$ and for $i \leq 2$, $\alpha_i$ is empirical set to 0 to get the best reconstruction quality.

The differences of our progressive reconstruction from the concept of deep supervision (Li et al., 2018; Wang et al., 2015) is, deep supervision adds auxiliary supervision signals to the intermediate network layers to facilitate the training convergence (L. Zhang et al., 2022), where the intermediate outputs are used to calculate the losses but not used to compose the final prediction. In contrast, 2D-to-3D reconstruction presents a different scenario. Firstly, all feature maps of 3DPX are trained to imitate the 3D reconstruction successively at different scales, instead of learning low-level or high-level semantic features representation. Secondly, applying progressive reconstruction guided by 3D ground truth on the intermediate outputs makes the feature maps into step-by-step reconstruction, where the intermediate reconstruction outputs (i.e., the feature maps) are directly used to facilitate the next step of reconstruction.

### 2) Hybrid MLP-CNN Block (HB)

The HB depicted in Fig. 2(d) combines the MLP block proposed by Tu et al (Tu et al., 2022) and convolutional layer together. Before MLP block, the first convolutional layer fuses the features that come from the former layer and the skip connection. In the MLP block, the multi-axis gated MLP layer (Tu et al., 2022) enables effective interactions between different feature spatial dimensions and capture both local and long-range dependencies of the input features. Then, following (Tu et al., 2022), a channel attention mechanism is used to weight the importance of different feature channels and improve the concentration on some channels while suppressing on others. The second convolutional layer perform the same operation with another skip connection to integrate the long-range attention information and recover the depth coherence. The output channel size of all these layers is 128. Max pooling layers and upsampling layers with a kernel size of 2×2 are used between adjacent blocks for downsampling and upsampling.

## C. Feature Projection Network (FPN) with Bidirectional Contrastive Multimodality Alignment (BCMA)

*1) 2D-3D bidirectional FPN without 3D labels*

Following (Hu et al., 2021), the cascaded FPN processes the 2D PX input and the reconstructed 3D structure in two parallel 2D/3D branches. It pyramidally projects features from the 3D branch to the 2D branch for feature fusion, enhancing the downstream classification and segmentation performance with the proposed BCMA (refer to Section III-C (2)). It is designed to exchange information between the PX images and the 3D synthetics to enhance the understanding in 2D domain. It acts as the function of skip connections between 2D and 3D sub-networks in the same encoder or decoder levels. While (Hu et al., 2021) dealt with real-world sparse volume, 3D reconstruction doesn't come with a 3D ground truth and is in dense representation. As a result, the parameters of 3D sub-network are manipulated by the gradient from 2D branch. Specifically, at each encoder or decoder level, the 2D features $f_{2D}f_{2D}^i$ with the shape of $H \times W \times C_{2D}$ and the 3D features $f_{3D}f_{3D}^i$ with the shape of $H \times W \times D \times C_{3D}$ is bidirectionally delivered to the opposite branch. On both branches, the features were first concatenated on the depth axis into $f_{2D+3D}$ with the shape of $H \times W \times (D + 1) \times C_{3D}$. Then the mixed 2D features $f_{2D}^{ii}$ is acquire by applying 1x1 average pooling on the depth axis while $f_{2D+3D}$ is directly used as the mixed 3D features for the next encoder or decoder to acquire $f_{2D}^{i+1}$ and $f_{3D}^{i+1}$. Hence, every convolutional layer in the 2D branch of the bidirectional FPN possesses not only the single view 2D features but also incorporates the 3D spatial features. As such, the FPN takes the advantages of the strength of both 2D images and its 3D reconstruction.

*2) Contrastive Multimodality Alignment (CMA)*

These features are then again bidirectionally delivered to the opposite branch or used as the input of contrastive multimodality alignment module, which was set after the second block of the 2D-3D joint encoder or decoder. Here, we introduce the formulation of CMA depicted in Fig. 2(b) and illustrate its difference from the contrastive loss (Chen et al., 2020; Y. Zhang et al., 2022). For a minibatch of $N$ images of 2D PX $\{X_1, X_2, \ldots, X_N\}$, the corresponding synthetic 3D structures is denoted as $\{X_1^*, X_2^*, \ldots, X_N^*\}$. The pair of 2D and 3D input from the same patient $(X_i, X_i^*)$ is regarded as a positive pair in calculating the contrastive loss. Denoting $z = c(X)$ as the 2D branch output and $z^* = c(X^*)$ as the 3D branch output of bidirectional projection pyramid encoder, the contrastive loss can be formulated as

$$\mathcal{L}_{CMA} = -\sum_{i=1}^{N} \log \frac{\exp(z_i \cdot z_i^*)/\tau}{\sum_{k=1}^{N} \mathbb{I}_{k \neq i} \exp(z_i \cdot z_k^*)/\tau},$$

where $\tau$ is a temperature parameter and $\mathbb{I}$ is an indicator function that returns 1 when $k \neq i$. Conceptually, this multimodality contrastive penalty promotes the representation alignment of synthetic 3D structure and its corresponding 2D input within the latent space, while increasing the difference of 3D and 2D representations from different patients. The difference between $\mathcal{L}_{CMA}$ and the original contrastive loss (NT-Xent loss (Chen et al., 2020)) is $\mathcal{L}_{CMA}$ focuses on the alignment or divergence of features from different modalities of the patient instead of different augmentations of the image. The original contrastive loss was applied on the batch with 2N images from two types of stochastic data augmentations, both between images of different augmentations and within the images of the same augmentation methods.

## Experimental Setup

### A. Patient and Dataset

We used 464 CBCT patient scans and their paired projected PX images. Among them, 91 scans were from a dataset partially released by Cui et al. (Cui et al., 2022), and 373 were from our private dataset from the School & Hospital of Stomatology, Wuhan University (WHU), China. The CBCT dataset collected by Cui et al. were scanned in routine clinical care, where patient required dental treatments such as orthodontics, dental implants, or restoration. The CBCT scans obtained from WHU were pathologically confirmed to exhibit 4 types of OCLs: ameloblastoma, dentigerous cyst, odontogenic keratocyst and radicular cyst. For CBCT dataset collected by Cui et al., the original resolutions were $400 \times 400$, with a varying height of approximately 280 pixels, at an interslice distance of $0.4\ mm \times 0.4\ mm \times 0.4\ mm$. For the WHU dataset, the original resolutions were $512 \times 512$, with a varying height of approximately 512 pixels, at an interslice distance of $0.3\ mm \times 0.3\ mm \times 0.3\ mm$. The WHU dataset contains annotations of the lesion area, acquired through a semi-automatic process and manual verified by experienced specialists.

### B. Preprocessing and Augmentation

To perform the 2D-to-3D reconstruction and 2D-3D joint analysis, paired PX image and its corresponding unfolded 3D structure are acquired. As PX image and CBCT pairs captured at the same time are rare, we obtained such pairs by projecting the PX and resampling the unfolded 3D structure on the CBCT based on the dental arch curves. The dental arch curves were hand marked on CBCT under the guidance of an experienced dental surgeon from the discipline of Oral and Maxillofacial Radiology, University of Sydney, Australia. The PX and its 2D segmentation mask was obtained by projecting on the CBCT and its 3D segmentation mask along the dental arch trajectory perpendicularly. The projection was conducted with a 0.2 mm unit size, encompassing a depth range of 40 mm and a height of 100 mm, with the width matching the length of the arch curve, typically around 200 mm. This projection region was simultaneously reformatted into an unfolded 3D structure using curved planar reformat method.

To evaluate the performance of the proposed 2D-3D joint analysis method, we augmented the projection process to obtain PX images with 4 types of misaligned filming error. These augmented data was employed to perform binary classification (2-class) with true and false labels, as well as multiclass classification (5-class) with 5 labels consists of regular, rotation-left, rotation-right, chin-up and chin-down. Following (Kwon et al., 2023), the misalignment of head was simulated by vertically and laterally rotating CBCT scans while fixing the position of dental arch trajectory. Considering the angular misalignment commonly encountered in clinical practice, we applied lateral rotation of 5 and 10 degrees to simulate head turn to the right and left, and vertical rotation of 5 degrees to simulate chin-up and chin-down positions. In total there were ~6 misaligned images generated for each case. As a results, the binary classification dataset presented an imbalanced data distribution (~1/7 images were with regular angle and ~6/7 with angular misalignment). For vertical rotation, the rotation center was set at the origin of the CBCT volume. For lateral rotation, the rotation center was located 10 mm below the teeth crowns and 15 mm posterior from the midpoint between the lower incisor when the head was correctly positioned. Following this, we obtained 2922 PX images with dimensions of [128, 256] and corresponding unfolded 3D structures with dimensions of [128, 256, 128].

### C. Comparison Methods

#### 1) 2D-to-3D Reconstruction

For 2D-to-3D reconstruction task, our 3DPX PRG backbone was compared to U-net and Residual CNN customized to fit 2D-to-3D reconstruction scenario, the state-of-the-art Oral3D, transformer-based method UNETR (Hatamizadeh et al., 2022), HB-based U-net, HB-based Residual CNN and their GAN models as competing methods. The channel numbers of the decoder branch in the existing methods were customized to fit the 2D-to-3D reconstruction target. The GAN training strategy on 3DPX is also reported.

#### 2) Angular Misalignment Classification

For angular misalignment classification task, our 3DPX was compared to several representative and recent end-to-end classification networks: (1) ResNet (He et al., 2016) for general 2D images; (2) xViTCOS (Mondal et al., 2022) – a visual transformer based method for chest radiography classification and, (3) Oral3D+ResNet (Song et al., 2021) – an enhanced 2D-3D joint network which analyzed the 2D PX and 3D structure reconstructed by Oral3D with a double branch customized ResNet. The customized ResNet has a 2D branch and a 3D branch, where the 3D features was bidirectionally exchanged with the 2D features by concatenation. We also reported the classification result guided by the ground-truth 3D structures (3DPX-real), as an upper bound result under our experimental setting of this tasks.

#### 3) OCLs segmentation

For OCLs segmentation task, the 3DPX was compared to the following well-established 2D segmentation models: (1) U-Net (Ronneberger et al., 2015) – a notable U-shape encoder-decoder structure designed for medical images; (2) Pyramid Vision Transformer (PViT) (Wang et al., 2022) – combining self-attention mechanism of transformer with pyramid feature extraction; (3) FCNT (Sanderson and Matuszewski, 2022) – combining the PViT with a Fully Convolutional Branch (FCB) for polyp segmentation on colonoscopy images, and (4) FCB of the FCNT. In addition, the 3DPX was further compared to a customized 2D-3D joint analysis methods: (5) Oral3D-FCB – modified FCNT to increase a 3D branch for feature fusion and using the Oral3D synthetics to refine the segmentation results. (6) 3DPX-real was compared as an upper bound result which was guided by the real 3D structures.

*4) 2D-3D feature fusion*

We compared the widespread fusion strategies with the adopted BCMA in our 3DPX. The compared early feature fusion strategies (applied to the input) were: (1) 2D-3D mean fusion for 2D ResNet and (2) 2D-3D stack fusion for 3D ResNet. The compared middle fusion strategies (applied to the output of every encoder) were: (1) unidirectional feature flow from 3D to 2D branch and fusion by concatenation, (2) unidirectional feature flow from 3D to 2D branch and fusion by mean operation, (3) unidirectional feature flow from 3D to 2D branch and fusion by max operation, (4) bidirectional feature exchange of 3D and 2D branch and fusion by max operation and (5) bidirectional feature exchange between 3D to 2D branch and fusion by mean operation (BPNet). The compared late fusion strategy (applied to the features before the classifier) was class activation map (CAM) concatenation.

### D. Evaluation Metrics

2D-to-3D reconstruction task was evaluated with Peak signal-to-noise ratio (PSNR), Structure similarity index (SSIM) and Dice similarity coefficient (DSC), consistent to (Song et al., 2021). PSNR was used to measure the difference between two signals to assess the density recovery of the reconstructed 3D structure. SSIM is a quality assessment metric that emphasize the differences between two images in brightness, contrast, and structure, where it takes into account of how well the local patterns in one image match the corresponding patterns in another images. DSC is commonly used to evaluate the similarity between the predicted and the ground-truth segmentation mask where a higher DSC score indicates better performance. In the assessment of 2D-to-3D reconstruction, we employed DSC as a metric to assess the structure deformation between 3D volume representation of ground-truth and reconstructed results. A threshold was applied to both the synthetic and ground-truth representation to obtain the high-density regions, mostly jaw bones and teeth. Then the extracted areas were viewed as segmentation masks on which the DSC scores were calculated. During the experiments, we set the threshold to 1.5 times the mean density.

The angular misalignment classification task was evaluated using precision, recall, accuracy, and F1 score. The lesion segmentation task was assessed also by DSC, Intersection over Union (IoU), precision and recall.

### E. Implementation Details

3DPX was implemented using Pytorch on 48GB NVIDIA A6000 GPU. Experiments were trained using Adam optimizer with a batch size of 8. The PGR backbone was trained with an initial learning rate of $4e^{-4}$, which decreased by half every 5000 iterations. The 2D-3D joint networks were trained with an initial learning rate of $1e^{-4}$. Training of joint classification was updated by cosine annealing schedular with a maximum iteration number of 200. Our implementation code is publicly available[1].

TABLE I
The comparison results between our 3DPX and existing 2D-to-3D reconstruction models. Methods marked with ∼ were customized from existing methods and/or with our proposed innovations to fit the 2D-to-3D reconstruction scenario. Best results are **bolded** and second-best underlined.

| Architecture | U-Net based | | | Residual CNN based | | |
|---|---|---|---|---|---|---|
| | PSNR | DSC | SSIM | PSNR | DSC | SSIM |
| Transformer (∼UNETR) | 14.76 | 60.57 | 60.3 | - | - | - |
| CNN (∼U-Net) | 14.76 | 62.22 | 67.72 | 15.21 | 61.17 | 70.58 |
| CNN GAN (Oral-3D) | 14.69 | 62.61 | 68.97 | 15.26 | 60.75 | 68.47 |
| Hybrid MLP-CNN (∼) | 14.99 | 62.2 | 68.55 | 15.42 | <u>61.64</u> | <u>72.25</u> |
| Hybrid MLP-CNN GAN (∼) | 15.11 | <u>63.7</u> | 68.42 | 15.23 | 60.96 | 69.02 |
| Progressive Hybrid MLP-CNN GAN(∼) | <u>15.51</u> | 63.21 | <u>72.17</u> | <u>15.45</u> | 60.69 | 71.22 |
| Progressive Hybrid MLP-CNN (3DPX) | **15.84** | **63.72** | **74.09** | **15.73** | **62.01** | **73.45** |

## Results

### A. 2D-3D Reconstruction

Table I presents the results of the reconstruction backbone of 3DPX compared with existing 2D-to-3D models. Pure CNN-based customized U-Net achieved SSIM of 67.72% and DSC of 62.22% on bone segmentation. The transformer-based U-Net (UNETR), however, deteriorated the reconstruction quality compared to the customized U-Net, resulting in a decrease of 2% in DSC and more than 7% in SSIM. Compared to the U-Net, Residual CNN exceled in retaining the gradient throughout convolution blocks due to its residual connection and therefore achieved a higher SSIM of 70.58%. The integration of GAN training strategy elevated the SSIM to 68.97% for the U-Net but decreased it to 68.47% for the Residual CNN in Oral-3D. The introduction of the Hybrid MLP-CNN Blocks (HB) enhanced both the basic models, notably increasing the SSIM for the Residual CNN to 72.25% and the DSC to 61.64%. The difference of improvement between HB U-net and HB Residual CNN was impacted by the ability of depth information restoration. The HB introduces both the expansion of the receptive field and a detrimental effect on depth coherence. Without progressive guidance, only a 2-step skip connection provided the depth restoration for HB U-Net. However, for HB Residual CNN, both the skip and the residual connections helped to maintain depth coherence. Despite this, the GAN strategy failed to deliver improvement with the presence of HB, both with and without progressive guidance. With progressive intermediate guidance, 3DPX outperformed the U-Net by over 6% in terms of SSIM, 7% in terms of DSC, and 1.1% in PSNR. In contrast, we only observed a moderate improvement with Residual CNN, where SSIM was increased by 2%. When progressive guidance sufficiently restored intermediate depth coherence in 3DPX with U-Net structure, the 3DPX with Residual CNN structure lost its efficacy and failed to surpass the more straightforward designed counterpart.

TABLE II
A breakdown ablative study of the reconstruction backbone and its impact on downstream classification task.

| Method | Reconstruction | | | 2 classes | | 5 classes | |
|---|---|---|---|---|---|---|---|
| | PSNR | DSC | SSIM | Acc (%) | F1 score | Acc (%) | F1 score |
| CNN (Baseline) | 14.76 | 62.22 | 67.72 | 84.4 | 0.755 | 85.3 | 0.845 |
| CNN + MLP (HB) | 14.99 | 62.2 | 68.55 (+0.83) | 86.2 (+1.2) | 0.786 | 91.6 (+6.3) | 0.905 |
| CNN + MLP + PGR (**Ours**) | **15.84** | **63.72** | **74.09 (+6.37)** | **92.4 (+8.0)** | **0.891** | **93.3 (+8.0)** | **0.923** |
| Oral3D | 14.69 | 62.61 | 68.97 (+1.25) | 88.4 (+4.0) | 0.833 | 92.9 (+7.6) | 0.925 |

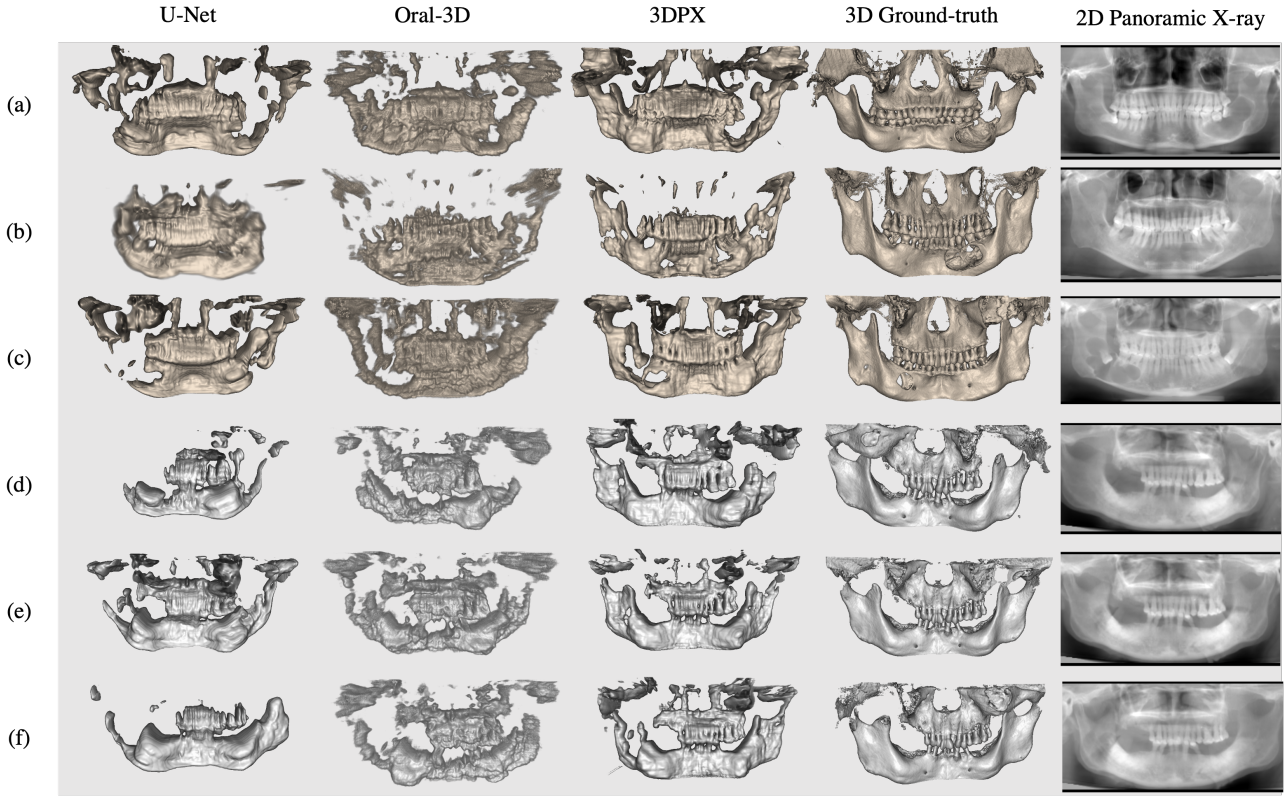

**Fig. 3.** Volume rendering of the reconstructed 3D structure of customized U-Net, Oral-3D, 3DPX and the ground-truth data (from left to right). (a-c) presents 3 cases with OCLs. Two types of misalignment augmentation are depicted, (a) regular PX capturing angle, (b) PX with 10 degrees of left rotation misalignment, and (c) PX with 10 degrees of right rotation misalignment.

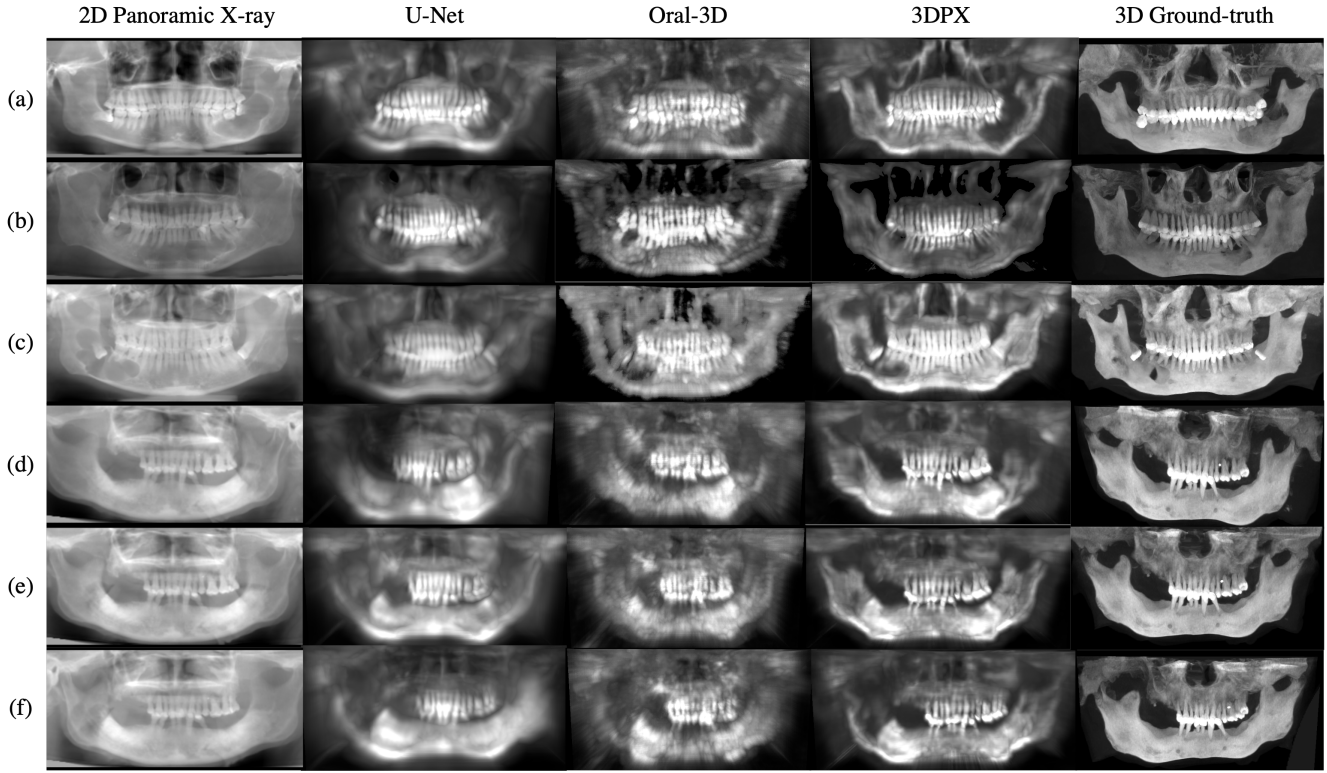

**Fig. 4.** MIP rendering of the reconstructed 3D structure against two comparison methods (customized U-Net and Oral-3D). (a-c) presents 3 cases with OCLs. Two types of misalignment augmentation are depicted, (a) regular PX capturing angle, (b) PX with 10 degrees of left rotation misalignment, and (c) PX with 10 degrees of right rotation misalignment.

A breakdown ablation study of the 3DPX reconstruction backbone on downstream classification tasks is shown in Table II. The improvements made by the HB and the PRG are reported step by step. Both of the modules contributed to the reconstruction quality and the classification results. During 3D reconstruction, PRG contributed more to the improvement compared to the HB. The SSIM increased by 0.83 with HB, while the combined effect with PRG further elevated it by 6.37. This trend persisted in binary classification, with PRG contributing more (4.8%) to accuracy compared to HB (1.2%). However, this effect did not reproduce on 5-category PX classification task. Minor improvement of reconstruction quality (+0.83) brought about relatively large boost on classification accuracy (+6.3). It reveals that the effectiveness of HB varies in different scenarios, and this will be further discussed later. Overall, the reconstruction backbone of our 3DPX surpassed Oral3D in all three scenarios, and greatly overrun it on reconstruction quality and binary classification which was more challenging. All the results on downstream experiments were from using the BPNet (Hu et al., 2021).

Fig. 3 illustrates threshold-based volume rendering of the 3D synthetic structures reconstructed by Oral-3D and 3DPX. The regions highlighted by the black box highlights that the 3DPX generated relatively fine-grained details for the anterior teeth and tooth implants. Additionally, it produced clearer expansions of one side of the jawbone ramus in the blue box and the other side of the jawbone body in the green box, which were caused by rotation misalignment during PX capturing. Although parts of the anatomy are missing for both methods, the important tooth sections are well preserved with 3DPX. Besides the threshold-based volume rendering which renders the high-density region with solid surface while erasing the low-density area, we also illustrate maximum intensity projection (MIP) of the synthesized volume in Fig. 4, to provide a global visualization from the coronal plane and quantitative comparison. The column 1 in Fig. 4 is the projected PX, the input of the 2D-to-3D reconstruction. Compared to 3DPX, U-Net failed to reconstruct major jawbone structures and Oral-3D was less capable of generating detailed tooth structure and clear boundary between bone and soft tissue and produced incoherent anatomical structures, especially apparent in the depth channels.

## B. 3D-guided PX Classification

TABLE IV
The comparison results of the proposed 3DPX and existing segmentation methods on OCLs segmentation.

| Method | Modality | DSC (%) | IoU (%) | Precision | Recall |
|---|---|---|---|---|---|
| Unet | 2D | 42.3 | 30.1 | 0.379 | 0.552 |
| FCB | 2D | 49.0 | 36.1 | 0.536 | 0.513 |
| PViT | 2D | 48.2 | 35.8 | 0.61 | 0.442 |
| FCNT | 2D | 47.8 | 34.6 | 0.466 | **0.577** |
| Oral3D+FCNT | 2D+3D | 50.1 | 37.7 | 0.639 | 0.461 |
| 3DPX (Ours) | 2D+3D | 55.2 | **43.7** | 0.643 | 0.539 |

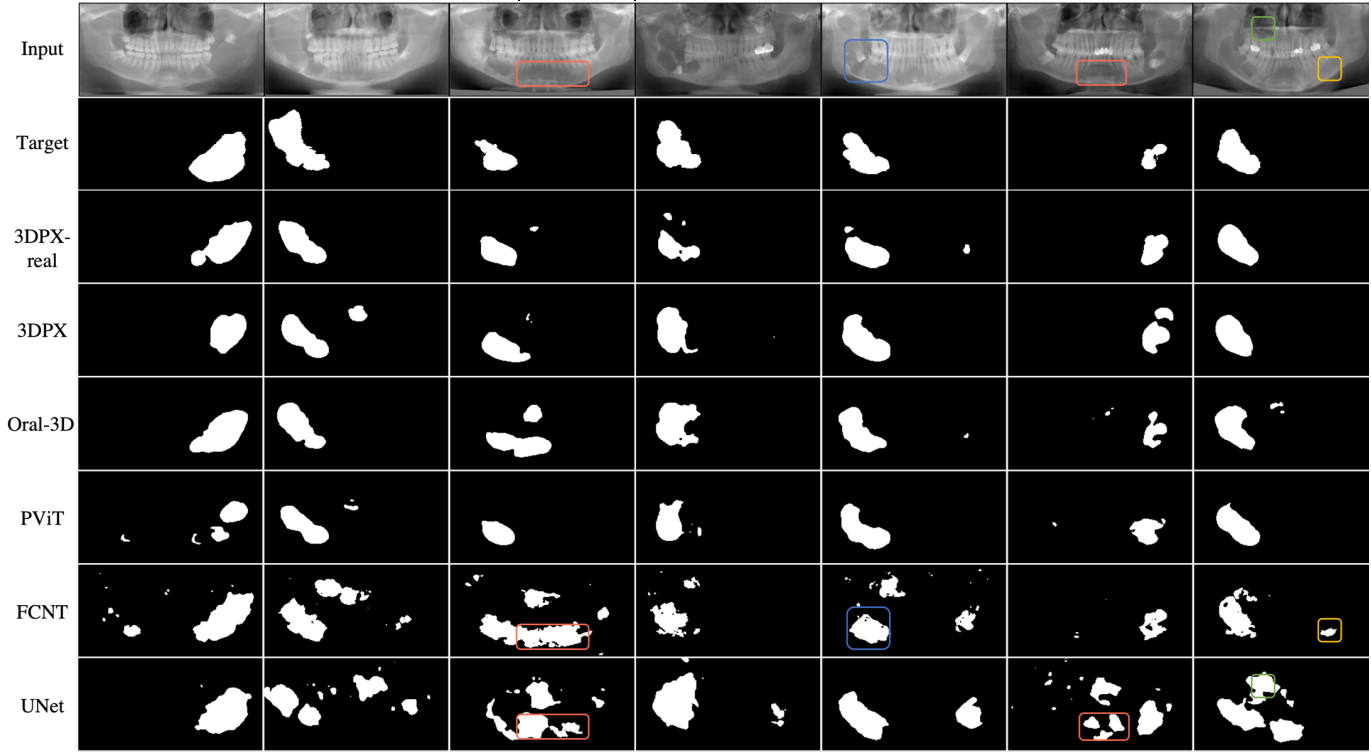

**Fig. 5.** Example cases of OCLs segmentation dataset and the predictions of 3DPX and considered existing architectures.

On the downstream binary and 5-class angular misalignment classification task, 3DPX is compared with representative CNN- and transformer-based 2D architectures as well as 2D-3D joint models with competing reconstruction backbone. The results are shown in Table III. On both tasks, 3DPX achieved highest results (93.3% and 94.2% overall accuracy). ResNet achieved an overall accuracy at 86.2% for binary classification and 88.9% for 5-class classification, setting the baseline for comparison. Overall, xViTCOS produced lower accuracy than ResNet (78.8% and 81.8%) when trained from scratch. When its transformer backbone was pre-trained on ImageNet, the accuracy was significantly boosted on both tasks to 92.4% and 92.9%. 3DPX-real produced an accuracy of 94.7% and 94.2% and F1 score of 0.942 and 0.948, enhanced from real 3D structure extracted from CBCTs. In comparison, experiment enhanced with Oral3D synthetics performed equally well with xViTCOS on multiclass classification, but the accuracy of 92.9% was 3 points lower from the upper limit of 95.6% using real 3D structures. On binary classification, Oral3D synthetics failed to compete with pre-trained xViTCOS with an accuracy of 88.4%. Instead, 3DPX surpassed both pre-trained xViTCOS and Oral3D reconstruction backbone with an accuracy of 93.3% on the binary task, and that of 94.2% on the multiclass task. It also prevailed on precision, recall and F1-score on both tasks, manifesting its robust proficiency on class-imbalanced dataset. Relative to the upper bound results, there remains room for improvement in 3DPX guided with synthetics, with a shortfall of 1.4% in binary accuracy and 0.8% in multiclass accuracy.

## C. 3D-guided PX Lesion Segmentation

TABLE V
Comparison results of the 2D-3D fusion strategy reported on binary angular misalignment classification task. Here Bi-di represents bidirectional feature projection, while Uni-di represents unidirectional feature projection from 3D to 2D branch.

| Strategy | Method | Accuracy (%) | Precision | Recall | F1 score |
|---|---|---|---|---|---|
| 2D only | ResNet18 | 86.2 | 0.802 | 0.744 | 0.767 |
| Early fusion | 3D-to-2D Mean | 79.6 | 0.398 | 0.5 | 0.443 |
|  | 2D-to-3D Tile | 89.3 | 0.834 | 0.844 | 0.839 |
| Middle fusion | Uni-di Cat | 87.6 | 0.811 | 0.801 | 0.806 |
|  | Uni-di Max | 92.4 | 0.87 | 0.92 | 0.891 |
|  | Uni-di Mean | 92.4 | 0.878 | 0.896 | 0.887 |
|  | Bi-di Max | <u>92.9</u> | <u>0.891</u> | 0.891 | <u>0.891</u> |
|  | BPNet | 92.4 | 0.87 | **0.92** | 0.891 |
| Late fusion | CAM Cat | 88 | 0.817 | 0.812 | 0.814 |
|  | BCMA | **93.3** | **0.891** | <u>0.91</u> | **0.9** |

We further examined the benefit of the reconstructed 3D structure on 2D analysis with OCLs segmentation task for PX. Table IV presents the comparison results of 3D-guided methods and existing 2D segmentation models for X-ray segmentation with respect to the mean DSC, IoU, precision and recall. All models were trained from scratch to ensure fairness. The overall performance of the 3D-guided methods surpassed that of prevailing 2D segmentation tasks. U-Net is presented as a baseline with a DSC score at 42.3%. Among four 2D segmentation networks, all performed better compared to U-Net with FCB achieving the best DSC and IoU scores at 49.0% and 36.1%, respectively, and with competitive results in precision and recall. The 3D structures generated by Oral3D slightly improved the segmentation performance from 49.0% to 50.1% on DSC score and from 36.1% to 37.7% on IoU score. In comparison, 3DPX showed an advantage in segmenting OCLs by achieving a DSC score of 55.2%, which is an improvement of 6% compared with 2D-only analysis and by 5% compared with Oral3D reconstruction backbone. Although it produced only second-best recall at 0.539 and second-best precision at 0.643, it outperformed FCNT and the real-3D 3DPX with a highest IoU at 43.7%, indicating that 3DPX produced less false positive and false negative errors. It was noted that 3DPX and 3DPX-real yielded results with minor distinction with 3DPX resulting in 0.1% lower in DSC but higher in IoU score by 1.0%.

TABLE VI
A breakdown ablation study of 3DPX on all downstream tasks.

| Method | 2 classes | | 5 classes | | segmentation | |
|---|---|---|---|---|---|---|
|  | Acc (%) | F1 score | Acc (%) | F1 score | DSC (%) | IoU (%) |
| 2D | 86.2 | 0.767 | 88.9 | 0.883 | 42.3 | 30.1 |
| 2D-3D PGR | 92.4 (+6.2) | 0.891 | 93.3 (+4.4) | 0.923 | **55.2 (+12.9)** | **43.7** |
| 2D-3D PGR+BCMA | **93.3 (+7.1)** | **0.9** | **94.2 (+5.3)** | **0.934** | 52.3 (+10.0) | 40.3 |

Figure 5 illustrates some examples of the PX images with OCLs and their predicted segmentation masks. In this example, it is challenging to distinguish between the bone resorption caused by lesion and inflammation (blue boxes) to the dark area caused by lower bone density (red boxes), thinner structure (yellow boxes), or less structure overlap (green boxes). The lower bone density is a transformation effect caused by structures being stretched when capturing PX images. In comparison, methods with the aid of 3D reconstructions was able to locate lesion more accurately based on the 3D semantics of the structure.

## D. 2D-3D Feature Fusion

Table V presents the comparison results of our BCMA relative to multi-modality feature fusion methods in PX binary classification task. In general, BCMA outperformed existing methods of early, middle, or late fusion stages, with a variant of BPNet (Hu et al., 2021) (Bi-di Max) stood out among existing fusion methods. It was a bidirectional fusion strategy by calculating channel-wise maximum instead of channel-wise mean of BPNet. It enhanced the classification accuracy from 92.4% of BPNet to 92.9% as the second best in all counterparts. In comparison, BCMA achieved an accuracy of 93.3%, outperforming the state-of-the-art BPNet by 0.9% and the customized Bi-di Max fusion module by 0.4%.

Table VI showcases the periodic impact of PGR and BCMA to evaluate the components of the proposed 3DPX. The second row highlights a notable enhancement in performance across all downstream tasks due to the incorporation of synthetic 3D information. Specifically, additional 3D synthetic data led to improvements in accuracy by 6.2% for PX binary classification, 4.4% for multi-class classification, and a substantial 12.9% increase in DSC for lesion segmentation. Based on this, the combination of BCMA in the downstream prediction task further refined the classification results in both binary and multiclass tasks by 0.9%. However, it is noteworthy that this fusion strategy did not sustain its efficacy in lesion segmentation; instead, it detrimentally impacted the DSC score.

## Discussion

Our main findings are: (1) the 3DPX outperformed all comparison methods on the classification of PX angular misalignment filming error, and on the segmentation of oral lesion; (2) comparing with existing 3D reconstruction methods, our 2D-to-3D reconstruction backbone achieved better quality, and this leads to greater benefit on the downstream tasks and, (3) BCMA outperformed existing fusion strategies in 3D-guided 2D PX classification tasks.

In the comparison between the 3DPX and existing classification and segmentation methods, 3DPX has achieved the best performance on all three PX analysis tasks (Table III and IV). This shows that the 3DPX was able to better leverage the discriminative features from the 3D synthetic data compared to other methods focused on using the PX image alone. We attribute this to two following reasons: (i) the depth information restored by the reconstruction backbone that serves as information supplement and, (ii) the synergy that is exploited between the 2D and 3D features on the downstream joint learning. In comparison, the 2D methods are limited to estimating the depth spatial information based on the pixel values based on its width and height axes which results in errors caused by lack of depth information that can aid in understanding the complex spatial relationship of tissues and skeletal structures.

We further note that compared to other comparison 2D-3D methods (Oral3D+ResNet in Table III and Oral3D+FCNT in Table IV), our 3DPX still achieved higher performance. Firstly, 3DPX achieved best reconstruction quality and outperformed all comparison methods (Table I). We noted that the Hybrid MLP-CNN network outperformed pure CNN-, MLP- or transformer-based methods. This is likely since that HB harmonized the strength and mitigates the demerits of both CNN and MLP block. Compared to 3D CNN that only leverage part of the depth information within the 3D convolution kernels, 2D CNN was able to leverage all the depth information in the feature channels. This facilitates the coherence in the depth dimension, but its receptive fields on the height and width axes are still limited. MLP and transformer share the same drawback on 2D-to-3D reconstruction in that they decouple the spatial and channel dimensions and process them independently. This decoupling induces limitations in their capability to preserve coherence on the depth (i.e., channel) dimension. This brings about their strength which is the ability to capture long-range dependence with low computation complexity. By conjoining MLP and CNN layer (Fig. 2(d)), HB was more efficient with long-range spatial dependency and capable of preserve essential depth coherence in our experiments. Furthermore, we noticed that the improvement of HB and PGR on reconstruction quality resulted in different ratio of accuracy improvement on classification with different label distribution (Table II). On the more balanced 5-class task, HB increased its accuracy by a large margin (6.3%) with relatively smaller SSIM improvement (0.83%). We attribute this to the HB's ability in improving long-range dependency, and thus the enlarged receptive field to model large deformation caused by the misalignment of PX. Based on HB, the further large improvement made by PGR on the imbalanced 2-class task was attribute to the detailed low-to-high guidance provided by the intermediate progressive guidance, which is essential to more challenging task. Secondly, the proposed BCMA outperformed existing feature fusion methods in 3D-guided 2D PX classification tasks (Table V). We attribute this to the fact that the contrastive penalty implicitly caused shifts in feature representation of both the 3D synthetics and the PX images. It propelled the latent representation of the 3D synthetic to approach that of its corresponding PX. Additionally, it seeks to differentiates the 3D synthetic images from the non-homologous PX images by applying penalty that creates a distinct separation in their features representation. Compared to the stand-alone bidirectional projection strategy proposed in BPNet (Hu et al., 2021), BCMA constrained the semantic expressions of 3D synthetics and controlled the information of potential reconstruction errors, while improved the synergy between 2D and 3D modalities as shown in Table VI. We also noted that BCMA had greater impact on the classification tasks compared to the segmentation task. We suggest this advantage is from the proposed contrastive penalty function with its ability to leverage semantic information at an abstract level (Chen et al., 2020). This abstraction can facilitate the convergence and complementation of latent representation between 2D and 3D features. Therefore, the more abstract the feature level, the more suitable the concept is to apply. To this end, BCMA was applied to the features before the last output layer for the classification 2D-3D downstream network, as the last layer yields the most abstract features before the final prediction. We further noted that 2D-3D joint analysis guided by 3DPX synthetic data yielded results that were only marginally inferior to those obtained from real 3D structures (3DPX-real). This indicates that in the given three tasks, even though the quality of 3D reconstruction was not at the same level of real 3D images, the combination of PGR and BCMA of our 3DPX developed nearly same amount of the guidance that real 3D images can contribute.

Our results demonstrated that 3D reconstructions, when innovatively used as a guide to PX, can significantly improve multiple PX downstream task performance. However, we have identified several areas for further improvement. Our future study will investigate the generalization of our 3DPX to other 2D medical imaging analysis, e.g., non-dental X-ray images which also suffers from no depth information. We suggest that our method should generalize well on panoramic radiography of other part of body, e.g., skull and spine, as they share similar unwrapped or straightened reformation without many structural occlusions. However, it may face challenges when reconstructing body regions with detailed depth information, such as abdomen. Another extension of 3DPX is to further explore the 2D-3D reconstruction of medical images by leveraging the advances in Neural Radiance Field (NeRF) methods (Corona-Figueroa et al., 2022; Liu et al., 2024). NeRF methods normally require more than one 2D images to reconstruct the 3D structures, as multiple views provide diversity of appearances of the scene from different angle. Compared to general X-ray, which was projected on a fixed orientation, PX image is inherently projected with multi views, which aligns with the function of NeRF method. Moreover, multiple emerging fusion strategies, including cross-attention mechanism (Song et al., 2022), could be considered for the fusion of real and synthetic features.

With a single panoramic X-ray, 3DPX's reconstruction backbone attempts to simulate the back projection from a single orthogonal direction. This has an inherent limitation where the synthesized 3D reconstruction lacks depth information leading to artefacts in the sagittal plane. For instance, in a typical anatomy, the reconstructed dentition should exhibit protruding teeth. However, in some cases, the reconstruction results in the inverted teeth instead. It's a known challenge in areas where only a single image is acquired, and this challenge remains to be solved.

## Conclusion

This study introduced 3DPX – a novel approach for panoramic X-ray (PX) image analysis guided by 2D-to-3D reconstruction. By integrating a PRG backbone and a downstream 3D-guided 2D network with the BCMA module, 3DPX effectively addressed the challenges of integrating synthesized 3D information reconstructed from panoramic X-ray (PX) to guide multiple 2D image analysis tasks. Experimental results on two datasets comprising 464 studies demonstrate the superior performance of 3DPX compared to state-of-the-art methods for 2D-to-3D oral reconstruction, PX classification, and OCLs segmentation. It is to our knowledge that this is the first study to examine the impact of synthesized 3D data on 2D imaging analysis, and the first study of 2D-3D joint analysis in the realm of medical imaging analysis. Furthermore, as a flexible pipeline for PX analysis with a single 2D input, 3DPX exhibits broad applicability to general X-ray imaging analysis.